 \documentstyle[preprint,aps]{revtex}  
\begin{document}
\draft



\title{Single Particle and Collective Structure for Nuclei near $^{132}$Sn}
\author{Jing-ye Zhang$^{(1)}$, Yang Sun$^{(1)}$,
Mike Guidry$^{(1,2)}$, L.L. Riedinger$^{(1)}$, and
G.A. Lalazissis$^{(3)}$
}
\address{
$^{(1)}$Department of Physics and Astronomy, University of Tennessee,
Knoxville, Tennessee 37996 \\
$^{(2)}$Physics Division, Oak Ridge National Laboratory,
Oak Ridge, Tennessee 37831 \\
$^{(3)}$Physik-Department, Technische Universit\"at M\"unchen,
D-85747 Garching, Germany
}

\date{\today}
\maketitle

\begin{abstract}
A new Nilsson single-particle structure is proposed
for neutron-rich nuclei near
$^{132}$Sn.
In general, a large reduction in spin--orbit interaction is required
and the neutron $N = 82$ gap persists in the new set of parameters.
The ground state deformations for
several isotopic chains are studied with
this set
and compared with the results of the standard set and with measured ones.
Collective bands in two even--even,
neutron-rich nuclei are calculated
using the Projected Shell Model with the new set of parameters
and improved agreement with existing
data is found.
\end{abstract}



\narrowtext

The study of nuclei far from $\beta$-stability
is an important topic in nuclear physics \cite{Hansen.79},
both because of  their intrinsic importance to nuclear structure and
for their importance in astrophysics.
One substantial step along this line is to investigate the neutron
rich nuclei where data are available, such as those
around $^{132}$Sn. These nuclei are of unique interest
because $^{132}$Sn is expected to have doubly
closed shells and the levels in nearby nuclei are expected to be largely
of single-particle nature.
We note in this regard that the 3$^{-}$ state in $^{132}$Sn lies at
4.3512 MeV, much higher than in $^{208}$Pb (2.6146 MeV)\cite{iso}. Thus,
$^{132}$Sn may be expected to be an even better closed shell nucleus
than $^{208}$Pb. Therefore, as an approximation, the single particle
and hole states observed experimentally in the nearest odd-A nuclei
are assumed to be pure in structure, and can be directly used as
references to construct the theoretical single particle scheme for
this region \cite{san}.

For the $^{132}$Sn region,
some experimental
data are beginning to fill the gaps between the known
single-particle spectrum in the stability valley and the unknown spectrum at
the
drip line.
For example, the single-particle structure above and
below the neutron number $N=82$ gap can be found from decay data for
$^{133}$Sn\cite {hof} and $^{131}$Sn \cite {fog}.
However, near the proton
$Z=50$ gap, less is known;
only a few particle states
from $^{133}$Sb \cite {blo,san} and one hole state from $^{131}$In
\cite{lun,fog} have been reported.
This paper proposes to construct a new single-particle spectrum
utilizing available experimental data, and to investigate
the impact of this spectrum on
ground state properties and collective excitations in the
region near $^{132}$Sn.

The Nilsson model with the ``standard'' set of parameters
\cite {nil,tod} has been
quite  successful in describing the single-particle structure for stable
nuclei.
Two parameters,
${\kappa}$ and ${\mu}$, appear in the Nilsson potential
\begin{eqnarray}
V ~=~ -{\kappa\hbar\omega^{\pm}_{0}} ~~\lbrack 2 l_{t}\cdot s + {\mu}
(l_{t}^{2} -
\langle l_{t}^{2}\rangle_{N}) \rbrack
\\
{\hbar\omega^{\pm}_{0}} ~=~ {\hbar\omega_{0}}~~\lbrack 1 \pm (N-Z)/3A \rbrack
\end{eqnarray}
where ``$+$" stands for neutrons
and ``$-$" for protons,
${\hbar\omega_{0}}$
is the harmonic-oscillator parameter,
$s$ is the intrinsic nucleon spin,
and $l_{t}$ is the orbital angular momentum in the
stretched coordinate basis \cite{nil}.
However,
the experimental information for
nuclei  near $^{132}$Sn indicates that the standard parameter set
cannot be correct for very neutron-rich nuclei. Thus, an
adjustment of these parameters is necessary if one hopes to describe this
region
using  a Nilsson single-particle spectrum.

Let us first consider the neutron single-particle structure.
To reproduce the observed single-neutron particle levels
above the $N=82$ gap, one must reduce
the strength of the spin--orbit interaction for $N=5$
with $l = 1$ and $l = 3$
substantially from the standard value in the stability valley
because of the observed
smaller separation between $f_{7/2}$
and  $f_{5/2}$ levels, and between $p_{3/2}$ and $p_{1/2}$ orbitals.
On the other hand,
the pair of $h$-orbitals require a larger value of $\kappa$
to position them properly above the gap ($h_{9/2}$)
and below it ($h_{11/2}$). For other orbitals below the $N=82$ gap,
one also requires a
reduction in  ${\kappa}$.

It is crucial to notice that, from Eq.\ (1),
if $\mu\le 0.50$ the energy order of neutron orbitals is
$d_{5/2}$, $g_{7/2}$, $s_{1/2}$, $d_{3/2}$, while for
$\mu > 0.50$, the energy order is
$g_{7/2}$, $d_{5/2}$, $d_{3/2}$, $s_{1/2}$.
Therefore, the ${\mu}$ parameter
for the $s,d$ shells and for the $g$ shell {\em cannot} be the same
if the experimental
energy order of $g_{7/2}$, $d_{5/2}$, $s_{1/2}$, $d_{3/2}$ \cite{fog}
is to be reproduced; in particular, the $g_{7/2}$ orbital is required
by the data to be about 800 keV
below the $d_{5/2}$ orbital. 
The necessity of introducing the $l$-dependent Nilsson parameters has been 
pointed out in Ref. \cite{seo} too.
For proton single-particle levels near $Z=50$ the experimental
information is less extensive and a reduction of $\kappa (N=4,5)$ plus
an adjustment of $\mu (N= 3,4,5)$ are necessary to reproduce  the limited set
of
known particle and hole spectra. However, since at present we don't have
further experimental information about the size of Z = 50 gap, we still keep
the standard kappa value for the pair of g-orbitals.
In Table I, we summarize the adjusted proton and neutron
$\kappa$ and $\mu$ parameters for the N = 3, 4, and 5 shells that
best reproduce the available
data in this region.
According to the most probable spin and parity assignment for the
parent state in $^{134}$In, the neutron separation energy in
$^{133}$Sn, and the semiempirical Woods-Saxon potential adjusted
from the data near $^{208}$Pb \cite{hof},
the ${\nu i_{13/2}}$ orbital should lie
above  3 MeV in excitation. In our case, the $N=6$
shell has not been modified and lies
at 3.4 MeV relative to the ${\nu f_{7/2}}$ orbital,
in a position consistent with all available information.
Table II shows
the corresponding single-particle level scheme from the experimental
data, that from the new
Nilsson parameters,
and that derived from
the standard set
of parameters \cite {nil,tod},
for nuclei around $^{132}$Sn.
Obviously, the new set of parameters nicely reproduces
the data while the standard set does not.

Relativistic Mean Field (RMF) theory \cite{Ring.96} with
nonlinear self-interactions between mesons has
been used
in many studies
of low-energy phenomena in
nuclear structure.
A careful investigation of isospin dependence for the spin--orbit interaction
in neutron-rich nickel and tin isotopic chains
indicates that the  spin--orbit
potential is considerably reduced,
resulting in smaller energy splittings between
spin--orbit partners\cite{LVR.97a}.
The spin--orbit interaction, which is central to the preceding discussions,
arises naturally in
RMF theory as a result of the Dirac structure of the nucleons.
Thus, it is relevant to consider the relation of the new single-particle
Nilsson
spectrum to that of the RMF.

The single-particle levels from RMF calculations
with three typical interactions \cite{Ring.96,LVR.97a}
(NL1, NL3, NLSH) are listed
in Table II as well.
Without any special parameter adjustment for this particular region,
the RMF single-particle levels are found to be reasonably close to the data.
Furthermore, the results
from both our new ${\kappa, \mu}$ Nilsson parameter set
and the RMF indicate a smaller
separation between spin--orbit partners that originates from a reduction
of the spin--orbit interaction strength
(except for our special adjustment of the neutron $\kappa$ for
$N=5$ to separate the $h_{11/2}$ and $h_{9/2}$ orbitals for the reasons
mentioned above).
The $N = 82$ gap survives in the new set of Nilsson parameter, as is also
true
in the RMF \cite{Ring1} and in the FRDM \cite{mol} for these neutron-rich
nuclei.
Fig.\ 1 shows the experimental neutron level scheme around the 82 gap
as well as calculated Nilsson levels with the standard set and proposed set
of parameters and the onse from RMF calculation with NL3 parameters.
It is obvious that the new set of parameters reproduces well the
observed particle and hole state levels.

Next, we consider the impact of the new Nilsson parameters
on other physical quantities.
Fig.\ 2 shows deformations extracted from Potential Energy Surface calculations
for cadmium, tellurium and xenon isotopes
by the usual Nilsson--Strutinsky--BCS approach \cite{nil}, employing both the
new
and the standard ${\kappa}$ and ${\mu}$ parameter sets. For reference,
deformations
extracted from experimental quadrupole
moment values \cite{ram} by an empirical formula \cite{lob} are included
too.
Unfortunately, the experimental deformations so far observed
are mostly for stable nuclei and
hence can't be used to check the new set of parameters.
However, the overall
difference between calculated deformations from the two sets of
parameters is not large, implying
that global ground-state properties of even--even nuclei such as the
deformation are not very sensitive to the details of the
single-particle structure. Therefore, the validity of the new parameter set
should be further checked by comparing with
other quantities that provide a more stringent test of
the single-particle structure.

The new Nilsson parameters should represent a
better basis from which more sophisticated wave functions can be
constructed. Therefore, we may test the new Nilsson parameters
by employing them in calculations that have a direct
connection with measured collective spectra.
For this purpose, we shall employ the Projected Shell Model (PSM)
to calculate the yrast bands of two even--even nuclei,
$^{136}$Te and $^{142}$Xe, for which limited data are available.

The PSM \cite{HS.95}
is a spherical shell model truncated in a deformed
Nilsson--BCS basis.
This
truncation is highly efficient if the single-particle basis is realistic
because the quasiparticle basis already
contains most of the correlations \cite{HS.95}.
Therefore, the quality of the Nilsson single-particle states
is crucial for the PSM results.
Once a good single-particle basis is provided to the PSM,
energy levels and many-body wave functions can be obtained
through shell model diagonalization.
It has indeed been shown that the PSM can nicely describe
the collective bands in normally deformed \cite{HS.95},
superdeformed \cite{SD190}, and
transitional nuclei \cite{jingye}.
This further permits the matrix elements
for processes such as electromagnetic transitions and
direct capture to be calculated as well.

In PSM calculations of the low-spin states relevant here,
the projected multi-quasiparticle
states consist of 0- and 2-qp (2$\nu$ and 2$\pi$) states for even--even
nuclei,
typically with a dimension of 50.
For the single-particle space we use three major shells:
N = 4, 5, and 6 for neutrons and N = 3, 4, and 5 for
protons. The deformations in the Nilsson single-particle basis were
$\varepsilon_2=0.060$ and $\varepsilon_4=-0.004$ for $^{136}$Te, and
$\varepsilon_2=0.150$ and $\varepsilon_4=-0.012$ for $^{142}$Xe.
These values are consistent with those presented in Fig.\ 2. All
the states within one nucleus were obtained from the diagonalization
in this (projected) shell model basis with the same basis deformation.

Fig.\ 3 shows results from the PSM
calculation.
It is obvious that the calculations employing
the new set of Nilsson parameters
nicely reproduce the data, while those with the standard set
of Nilsson parameters determined in the stability valley are in much poorer
agreement.
More important  than the quantitative differences is that the
qualitative
nature of the collective excitations is very different for the two
parameter sets.  For example, in $^{142}$Xe,
the new set gives (correctly) a more rotational
structure,
while the older set suggests a more vibrational structure,
even though the calculated deformations are very similar.
Here it is the change in single particle states that gives rise
to the different nature of the yrast sequence.

Around the mass-140 region, especially near $Z=56$, octupole
degrees of freedom have been found to be important in earlier
calculations
\cite{egi,wit,mol}.
In the present PSM calculation, the octupole degree of freedom has not been
included. Therefore, the number of nuclei that can be used to examine the
proposed Nilsson parameter set has been restricted
to nuclei that are expected to have
little octupole coupling.
We note that the
separation between the two key orbitals that is responsible for the main
octupole
correlation in the mass region discussed,
${\nu i_{13/2}}$ and ${\nu f_{7/2}}$, is about 1 MeV
larger from  the new set of parameters than that from the standard one (see
table
II), which should result in a weaker octupole correlation. Therefore
we do not expect the inclusion of octupole forces to
alter our discussion of the examples shown here.
In the future, one should investigate quantitatively the influence of
octupole deformation on the yrast band spectra for this neutron-rich
mass region based on the new set of Nilsson parameters with larger separation
of
the ${\nu i_{13/2}}$ and ${\nu f_{7/2}}$ orbitals.

Because of the
limited present information on empirical single-particle spectra in this
region,
this new
set of parameters should be employed with confidence
only for neutron-rich nuclei with $Z = 46 - 56$
and $N = 68 - 96$.  Furthermore,  the present parameterization is on firmer
ground
for nuclei
with proton number beyond 50, since there is only one hole state known
experimentally below the $Z=50$ gap.
Nevertheless,
the evidence cited above suggests that a parameterization at least
similar to the
one presented here will be  required to describe the physics of this
entire region of neutron-rich nuclei.

In summary, a new set of Nilsson parameters is proposed for neutron-rich nuclei
around $^{132}$Sn.  These parameters differ substantially from those commonly
employed for stable nuclei, and nicely reproduce the existing
data.
Reduced separation between spin--orbit partners and persistence
of the neutron $N=82$ gap are found in the new Nilsson diagram.  We find many
similarities between the new single-particle scheme and results from
Relativistic Mean Field Theory.
Available data for
yrast bands in
even--even nuclei around $^{132}$Sn are well reproduced by
PSM calculations with the new set of Nilsson states
as a basis, but
the spectra are not even
qualitatively in agreement with data when the standard set is
used. However, the nuclei that can be used to examine the proposed
parameter set is quite limited at present,  and
it is obvious that more experimental data are needed to further
fix the Nilsson parameters for these neutron rich nuclei. Improvements
in the theoretical model, such as the inclusion of the octupole degree of
freedom
is required as well.

\acknowledgments
We thank Drs.\ C. Bingham,  J. Dobaczewski
and C. L. Wu for valuable discussions.
                            Research at the University of
                            Tennessee is supported by the U.~S. Department
                            of Energy through Contract No.\
                            DE--FG05--96ER40983.
Oak Ridge National Laboratory, is managed by Lockheed Martin
                        Energy Research Corp.\ for the U.~S.
                        Department of Energy under Contract No.\
                        DE--AC05--96OR22464.

\baselineskip = 14pt
\bibliographystyle{unsrt}

\begin{figure}
\caption{ Single neutron particle (a) and hole(b) states.
Experimental (left-most) and calculated results from Nilsson
model with new parameter set, from standard set,
and from RMF with NL3 parameterization.}
\label{figure.1}
\end{figure}

\begin{figure}
\caption{Ground state deformations: experimental values
\protect\cite{ram} (dots), absolute values from calculations
with new set of parameters
(open triangles), and with standard parameter set (open circles)
for Cd, Te and Xe isotopes.
}
\label{figure.2}
\end{figure}

\begin{figure}
\caption{Yrast bands in $^{136}$Te and $^{142}$Xe from
experimental data (dots) \protect\cite{iso} and
PSM calculations with the new set of Nilsson parameters (open
triangles),
and with
the standard Nilsson parameters (open circles).
}
\label{figure.3}
\end{figure}

\newpage
\begin{table}[h]
\begin{center}
\caption{Nilsson Parameters $\kappa$ and $\mu$ around $^{132}$Sn}
\vspace{0.5cm}
\begin{tabular}{c|c|c|c|c|c|c|c|c|c}
N&$\it l$&$\kappa_{p}$& new $\kappa_{p}$&$\mu_{p}$&new $\mu_{p}$
&$\kappa_{n}$& new $\kappa_{n}$&
$\mu_{n}$&new $\mu_{n}$\\\hline
3 &1,3&0.090& &0.300&0.340 &0.090& &0.250& \\
\hline
4 &0,2&0.065&0.039 &0.570&0.760 &0.070&0.039 &0.390&0.330 \\
  & 4 &     &0.065 &     &      &     & &     &1.089 \\
\hline
5 &1,3&0.060&0.052 &0.650&0.645 &0.062&0.035 &0.430&0.090 \\
  & 5 &     &      &     & &     &0.082 &     &0.490 \\
\end{tabular}
\end{center}
\end{table}

\begin{table}[h]
\begin{center}
\caption{Level Scheme of $^{131}$Sn (neutron hole),
$^{133}$Sn (neutron particle),
$^{131}$In (proton hole)
and $^{133}$Sb (proton particle)}
\vspace{0.5cm}
\begin{tabular}{c|c|c|c|c|c|c|c}
&orbital&exp.&new $\kappa$, $\mu$& old $\kappa$, $\mu$&NL1&NL3&NLSH\\
\hline
$\nu$- & $d_{3/2}$  & 0.000  & 0.000  & 0.000  &0.000 &0.000 &0.000 \\
hole   &$h_{11/2}$ & -0.242 & -0.265 & -0.044 &+2.019&+1.249&+0.446\\
&$s_{1/2}$  & -0.332 & -0.346 & -0.403 &+0.308&+0.423&+0.461\\
&$d_{5/2}$  & -1.655 & -1.695 & -3.047 &-1.623&-1.657&-1.759\\
&$g_{7/2}$  & -2.434 & -2.435 & -2.108 &-2.309&-3.403&-4.224\\\hline
$\nu$- & $f_{7/2}$  & 0.000 & 0.000 & 0.000 & 0.000 & 0.000 &0.000 \\
part.  & $p_{3/2}$  & 0.854 & 0.883 & 3.399 & 0.868&0.860&0.702\\
&$h_{9/2}$  & 1.561 & 1.481 & 0.681 & 1.956&0.956&0.369\\
&$p_{1/2}$  & 1.656 & 1.797 & 5.018&1.203&1.202&1.033 \\
&$f_{5/2}$  & 2.005 & 2.122 & 3.777 & 1.468&1.434&1.312 \\
&$i_{13/2}$ &       & 3.408 & 2.476 & 4.729&3.951&3.266\\
\hline
\hline
$\pi$- & $g_{9/2}$  & 0.000  & 0.000  & 0.000 &0.000 &0.000 &0.000  \\
hole   &$p_{1/2}$ & -0.363 & -0.358 & -1.102&-1.546&-1.010&-0.426\\\hline
$\pi$- &$g_{7/2}$  & 0.000  & 0.000  & 0.000 &0.000 &0.000 &0.000 \\
part.  &$d_{5/2}$  & 0.962 & 0.965 & 0.472&1.851&2.814&3.563\\
       &$d_{3/2}$  & 2.440 & 2.408 & 2.880&3.534&4.503&6.257\\
       &$h_{11/2}$  & 2.793 & 2.772 & 1.536&4.331&4.611&4.629\\
\end{tabular}
\end{center}
\end{table}

\end{document}